\documentclass[twocolumn,prl,preprintnumbers]{revtex4}

\usepackage{times}
\usepackage{graphicx}
\usepackage{dcolumn}
\usepackage{bm}

\begin{document}

\title{\huge Prediction and statistics of pseudoknots in RNA structures\\
  using exactly clustered stochastic simulations}

\author{A.~Xayaphoummine, T.~Bucher, F.~Thalmann \&
  H.~Isambert\footnote{Corresponding author: herve.isambert@curie.fr ~~New address: Institut Curie, CNRS-UMR168,  11 rue P \& M Curie,
  75005 Paris, France.}}

\vskip 0.2cm
 
\affiliation{Laboratoire de Dynamique des Fluides Complexes, CNRS-ULP, Institut de Physique, 3 rue de l'Universit\'e, 67000 Strasbourg, France}%

\maketitle

\noindent
{\bf 
Ab initio RNA secondary structure predictions have long dismissed
helices interior to loops, so-called  pseudoknots, despite their 
structural importance.
Here, we report that many pseudoknots can be predicted through long
time scales RNA folding simulations, which follow the stochastic
closing and opening of individual RNA helices.
The numerical efficacy of these stochastic simulations relies on an 
${\cal \bf O}(n^2)$ clustering algorithm which computes time averages 
over a continously updated set of $n$ reference structures. 
Applying this  {\em exact} stochastic clustering approach, we typically obtain 
a 5- to 100-fold simulation speed-up  for RNA sequences up to 400 bases, 
while the
effective acceleration can be as high as 10$^5$-fold for short multistable 
molecules ($\le$~150 bases).
We performed extensive folding statistics on random and natural RNA sequences, 
and found that pseudoknots are unevenly distributed amongst RNA
structures and account for up to 30\% of base pairs in G+C rich 
RNA sequences (Online RNA folding kinetics server including pseudoknots : 
{\sf http://kinefold.u-strasbg.fr/}).}

\small

\vspace{.2 cm}

The folding of RNA transcripts is driven by intramolecular GC/AU/GU base 
pair stacking interactions. This primarily leads to the formation 
of short double-stranded RNA helices connected by unpaired regions.
{Ab initio} RNA folding prediction restricted to {\it tree-like} secondary 
structures is now well 
established\cite{nussinov,waterman,nussinov2,zuker,mccaskill,turner,vienna,higgs} and has 
become an important 
tool to study and design RNA structures which remain by and large refractory 
to many crystallization techniques.
Yet, the accuracy of these predictions is difficult to assess
--despite the precision of stacking interaction tables\cite{turner}--
due to their a priori dismissal of pseudoknot helices, Fig~1A.

Pseudoknots are regular double-stranded helices which provide specific 
structural rigidity to the RNA molecule by connecting different 
``branches'' of 
its otherwise more flexible {\it tree-like} secondary structure (Figs~1A-B).
Many ribozymes, which require a well-defined 3D enzymatic shape,
have pseudoknots\cite{pleij,tinoco,westhof,williamson1,woodson1,williamson2,woodson2,herschlag,ferre}. 
Pseudoknots are also involved in mRNA-ribosome interactions
during translation initiation and frameshift regulation\cite{frameshift}.
Still, the overall prevalence of pseudoknots  has proved difficult to 
ascertain from the limited number of RNA structures known to date.
This has recently motivated several attempts 
to include pseudoknots in RNA secondary structure
predictions\cite{gultyaev,eddy,isambert}.

There are two main obstacles to include pseudoknots in RNA structures:
a structural modeling problem and a computational efficiency issue.
In the absence of data bases for pseudoknot energy parameters, 
their structural features have been modeled at various descriptive levels 
using polymer theory\cite{mironov,gultyaev,isambert}.
From a computational perspective, pseudoknots have proved
not easily amenable to classical polynomial minimization 
algorithms\cite{eddy} due to their intrinsic non-nested nature. 
Instead, simulating 
RNA folding dynamics has provided an alternative avenue
to predict pseudoknots\cite{mironov,isambert} in addition to bringing 
some unique insight into the kinetic aspects of RNA 
folding\cite{higgs,isambert}.

\begin{figure}
\includegraphics{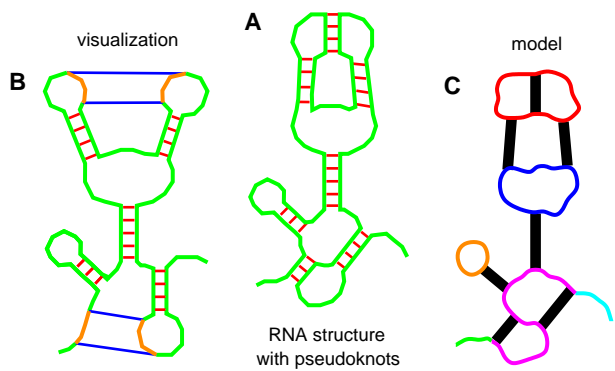}
\caption{\label{fig:wtgI} 
\small
 {\bf A} An RNA secondary structure with pseudoknots.
 {\bf B} Minimum set of helices {\it defined} as ``pseudoknots'' and 
visualized for convenience by colored 
single-stranded regions connected by two straight lines.
{\bf C} 
The entropic cost 
of the actual 3D structural constraints is evaluated by modeling
RNA helices as stiff rods (black) and single-stranded regions as ideal 
polymer springs. Colored single-stranded circuits {\it define} 
quasi-independent structural domains referred to as ``nets'' 
in ref\protect\cite{isambert}.}
\end{figure}

Yet, stochastic RNA folding simulations can become relatively
inefficient due to the occurrence of {\em short cycles} amongst 
closely related configurations\cite{mironov}, which typically differ
by a few helices only. 
Not surprisingly, similar numerical pitfalls have been recurrent 
in stochastic simulations of other trapped dynamical
systems\cite{frenkel,BKL,mezard,voter,pande}.

To address this computational efficiency issue and capture the slow 
folding dynamics of RNA molecules, we have developed 
a generic algorithm which greatly accelerates RNA folding stochastic 
simulations by  {\it exactly} clustering the main short cycles along the 
explored folding paths. The general approach, which may prove useful
to simulate other trapped dynamical systems, is discussed in the main
subsection of {\it Theory and Methods}.
In the {\it Results} section,
the efficacy of these {\it exactly clustered stochastic} (ECS) simulations 
is first compared
to non-clustered RNA folding simulations, 
before being used to predict the prevalence of  pseudoknots in RNA 
structures on the basis of the structural model introduced in 
ref\cite{isambert} and briefly reviewed hereafter.

\vspace{.5cm}

\noindent
{\large \bf Theory and Methods}
\vspace{.2cm}

\noindent
{\bf Modeling and visualizing pseudoknots in RNA structures.} 
We model the 3D constraints associated with pseudoknots using 
polymer theory.
The entropy costs of pseudoknots and internal, bulge and hairpin
loops are evaluated on the same basis by modeling the secondary
structure (including pseudoknots) as an assembly of stiff rods
--representing the helices-- connected by polymer springs
--corresponding to the unpaired regions, Fig~1C.
In practice, free energy computations involve the labelling of
RNA structures into constitutive ``nets'' --shown as 
colored circuits on Fig~1C-- to account for the stretching of the 
unpaired regions linking the extremities of pseudoknot helices, 
see ref\cite{isambert} for details. 
In addition, free energy contributions from base pair stackings,
terminal mismatches and co-axial stackings are taken from the
thermodynamic tables measured by the Turner lab\cite{turner}.

The main limitation of this structural model is the absence 
of hardcore interactions, which could stereochemically 
prohibit certain RNA structures with {\it either} long pseudoknots 
({\it e.g.}, $>$11bp, one helix turn) {\it or} a large proportion 
of pseudoknots ({\it e.g.}, $>$30\% of formed base pairs). However, we 
found that such  stereochemically improbable structures account for less than
1-to-10\% of 
all predicted structures, depending on G+C content (see Results section). 
Hence, in practice, neglecting hardcore interactions is rarely 
a stringent limitation, except for a few, somewhat pathological cases.

Although the presence of pseudoknots in an RNA structure is not 
associated to a unique
set of helices, it is
convenient for visualization and statistics purposes to
{\it define} the set of pseudoknots as the minimum set of helices
which should be imagined broken to obtain
a tree-like secondary structure,  Fig~1B. 
Finding such a minimum set
(with respect to the number of base pairs 
or their free energy) amounts to finding the maximum 
tree-like set amongst the formed helices and can be done in
polynomial time using a classical ``dynamic programming'' algorithm.

\vspace{0.3cm}
\noindent
{\bf Modeling RNA folding dynamics and straightforward stochastic algorithm.}
RNA folding kinetics is known to proceed through rare stochastic openings and 
closings of individual RNA helices\cite{porschke74bonnet98}. The time limiting 
step to transit between two structures sharing essentially all but one helix  
can be assigned Arrhenius-like rates,
$k_\pm = k^\circ \times \exp(-\Delta G_\pm/kT)$, where $kT$ is the thermal 
energy. 
$k^\circ$, which reflects only local stacking processes within a 
transient nucleation core, has been estimated
from experiments on 
isolated stem-loops\cite{porschke74bonnet98}($k^\circ \simeq 10^8$~s$^{-1}$), 
while the
free energy differences $\Delta G_\pm$ between the  transition states  and 
the current configurations (Fig~2) can be evaluated by combining the stacking
energy contributions and the global coarse-grained structural model described 
above, Fig~1C.

Simulating a stochastic RNA folding pathway amounts to following one 
particular  
stochastic trajectory within the large combinatorial space of mutually 
compatible helices\cite{mironov}.
Each transition in this discrete space of RNA structures corresponds to the 
opening {\it or} closing of a {\it single} helix, possibly followed by 
additional helix elongation and 
shrinkage rearrangements to reach the new structure's equilibrium compatible
with a minimum size constraint for each formed helix\cite{isambert} 
(base pair zipping/unzipping kinetics occurs on much shorter 
time scales than helix nucleation/dissociation). 
For a given RNA sequence, the total number of possible helices (which 
roughly scales as $L^2$, where $L$ is the sequence length) sets the 
local connectivity of the discrete structure space and therefore the number 
of possible transitions from each particular structure.

Formally, we consider the following generic model. 
Each structure or ``state'' $i$ is connected to a finite, 
yet possibly state-to-state varying number of neighboring 
configurations $j$ via transition rates $k_{ji}$ (the right-to-left matrix 
ordering of indices is adopted hereafter).
As $k_{ji}$ is the average number of transitions from state $i$ to state
$j$ per unit time, the lifetime $t_i$ of configuration $i$ corresponds 
to the average time before {\it any} transition towards a neighboring 
state $j$  occurs, i.e., $t_i = 1/{\sum_{\langle j \rangle}} k_{ji}$, 
and the transition 
probability from state $i$ to state $j$ is $p_{ji} = k_{ji}t_i$, with
${\sum_{\langle j \rangle}}p_{ji} = 1$, as expected, for all state $i$.

\begin{figure}
\includegraphics{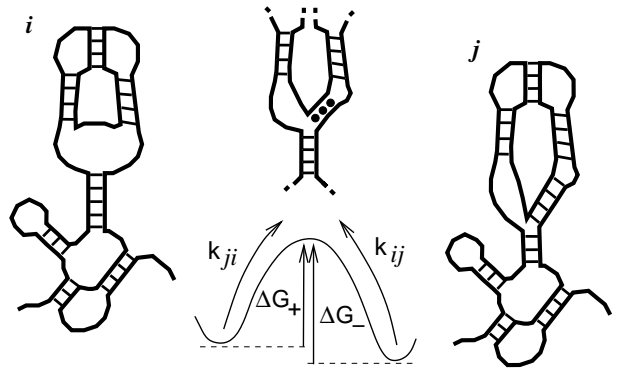}
\caption{\label{fig:wtgI} 
\small
Stochastic transitions over a thermodynamic barrier $\Delta G_\pm$ to 
close and open an individual helix between two neighbor RNA structures,
 $i$ and $j$. Nucleation of the new helix usually involve some  local
unzipping of nearby helices at the barrier and further base pair
rearrangements to reach equilibrium in the new structure 
$j$\protect\cite{isambert}.}
\end{figure}

Hence, in the straightforward stochastic algorithm\cite{mironov,isambert}, 
each new transition 
is picked at random with probability $p_{ji}$ while the effective time is 
incremented with the lifetime $t_i$ of the current configuration 
$i$\cite{distribution}.
However, as mentioned in the introduction, the efficiency of this approach 
is often severely impeded by the existence of {\it kinetic traps} 
consisting of rapidly exchanging states. 

\vspace{0.3cm}
\noindent
{\bf Exactly clustered stochastic (ECS) simulations.}
As in the case of RNA folding dynamics, the simulation of other trapped 
dynamical systems generally presents a computational efficiency issue.
In particular, 
powerful numerical schemes  have been developed to compute the elementary 
escape times from traps for a variety of 
simulation techniques\cite{frenkel,BKL,mezard,voter,pande}.
Still a pervasive problem usually remains for most applications due 
to the occurrence of {\em short cycles} amongst trapped states, and
{\it heuristic} clustering approaches have been proposed to overcome these
``numerical traps''\cite{krauth}.

To capture the slow folding dynamics of RNA 
molecules, we have developed  an {\it exact} stochastic algorithm 
which accelerates the simulation by numerically integrating
the main short cycles amongst trapped states.
This approach being quite general, it could prove useful to simulate other 
small, trapped dynamical systems with coarse-grained degrees of freedom. 

In a nutshell,
the ECS algorithm aims at overcoming the numerical pitfalls 
of kinetic traps by ``clustering'' some recently explored configurations 
into a single, yet continuously updated cluster $A$ of $n$ reference states. 
These clustered configurations are then 
{\it collectively} revisited in the subsequent stochastic 
exploration of states.
Although stochasticity is ``lost'' for the individual clustered
states, its
statistical properties are, however, {\it exactly} transposed at the 
scale of the set $A$ of the $n$ reference states.
This is achieved as follows. For each {\it pathway} $C^A_m$ on $A$, a 
{\it statistical weight} $W^{C^A_m}=\prod^{C^A_m} p_{lk}$ is defined,
where  $k$ and $l$ run over all {\it consecutive} states along $C^A_m$ 
from its ``starting'' state $i$ to its ``exiting'' state $j$ on $A$.
The $n\!\times\!n$ 
probability matrix $P^A$ which {\it sums} the statistical weights 
$W^{C^A_m}$ over {\it all pathways $C^A_m$ on $A$} between any 
two states $i$ and $j$ of $A$ is then introduced,
\begin{equation}
P^A_{ji}= \sum^{C^A}_{m:j \leftarrow i} W^{C^A_m}= \sum^{C^A}_{m:j \leftarrow
  i} \biggl(\ \prod_{j \leftarrow i}^{C^A_m} p_{lk} \biggr), \label{proba}
\end{equation}
and the exit probability 
to make a transition {\it outside} $A$ from the state $j$ is noted:
$p^{eA}_j = 1 - {\sum^A_{\langle k \rangle}} p_kj$. 
Hence, starting from state $i$, the probability to exit the set $A$ at 
state $j$ is $p^{eA}_j P^A_{ji}$, with
$\sum^A_j p^{eA}_j P^A_{ji} = 1$, for all $i$ of $A$.

Thus, in the ECS algorithm, one {\it first} chooses
at random with probability $p^{eA}_j P^A_{ji}$ the reference state $j$
of  $A$ from which a new transition towards a state $k$ {\it outside} $A$ will
{\it then} be chosen stochastically with probability $p_{kj}/p^{eA}_j$. 
Meanwhile, the physical quantities of interest, like the cumulative time 
lapse ${t}^A_{ji}$ to exit the set $A$ from $j$ starting at $i$, are 
{\it exactly} averaged over {\it all} (future) pathways 
from $i$ to $j$ within $A$,
as explained in the next subsection.
{\it Finally}, the new state $k$ is added to the reference set $A$ 
whilst another reference 
state is removed, so as to update $A$, as discussed in 
{\it The ${\cal O}(n^2)$ algorithm} subsection. 

\vspace{0.3cm}
\noindent
{\bf Exact averaging over all future pathways.}
We start the discussion with the {\it path average} time lapse 
to exit the set $A$.
Let us introduce the time lapse transform of ${P}^A$: ${\cal T}[{P}^A]\{t\}=\tilde{P}^A\{t\}$, which {\it sums} 
the {\it weighted cumulative lifetimes} 
$\bigl(\sum^{C^A_m}t_h\bigr)\prod^{C^A_m} p_{lk}$ over 
{\it all pathways $C^A_m$ on $A$} between any two states $i$ and $j$ of $A$,
\begin{equation}
{\cal T}[{P}^A]_{ji}\{t\}=\tilde{P}^A_{ji}\{t\}= \sum^{C^A}_{m:j \leftarrow i} \Biggl[ \biggl(\sum_{j \leftarrow i}^{C^A_m}t_h\biggr)\prod_{j \leftarrow i}^{C^A_m} p_{lk} \Biggr], \label{time}
\end{equation}
where the $t_h$'s are summed over all {\it consecutive} states $h$
--from $i$ to $j$ {\it included}-- along each pathway $C^A_m$.
Hence, the mean time $\bar{t}^A_i$ to exit $A$ from {\it any state} $j$ of $A$ 
starting from configuration $i$ is,
$\bar{t}^A_i =\sum^A_j p^{eA}_j  \tilde{P}^A_{ji}\{t\}$.
However, in the context of the ECS algorithm, the time lapse of 
interest is $\bar{t}^A_{ji}$, the mean time to exit $A$ from a 
{\it particular} state $j$,
$\bar{t}^A_{ji} = {p^{eA}_j  \tilde{P}^A_{ji}\{t\} / p^{eA}_j P^A_{ji}} =  {\tilde{P}^A_{ji}\{t\} / P^A_{ji}}$.

The average of any 
path cumulative quantity of interest $x_i$ can be similarly obtained 
by introducing the appropriate $\tilde{P}^A\{x\}$ matrix.
In particular,
the instantaneous efficiency of the algorithm is well reflected by the 
average pathway {\it length} $\bar{\ell}^A_{ji}$ between any two states 
of $A$,
\begin{equation}
\bar{\ell}^A_{ji} = {\tilde{P}^A_{ji}\{\ell\} / P^A_{ji}}, \label{meanlengthji}
\end{equation}
where $\tilde{P}^A_{ji}\{\ell\}= \sum^{C^A}_{m:j \leftarrow i} \bigl[
  \bigl(\sum^{C^A_m} 1\bigr)\prod^{C^A_m} p_{lk} \bigr]$, 
with $\sum^{C^A_m} 1$ corresponding to the length of the pathway $C^A_m$ 
(1 is added at each state  along each 
pathway $C^A_m$). Hence, starting from state $i$, $\bar{\ell}^A_{ji}$ 
corresponds to the {\it average number of transitions} that {\it would} have 
to be performed by the straightforward algorithm before exiting the set $A$ 
at state $j$.
As expected, $\bar{\ell}^A_{ji}$ can be very large for a trapped dynamical 
system,
which accounts for the efficiency of the present 
algorithm. Since the approach is {\it exact}, there is, 
however, no {\it a priori} requirement on the trapping condition of the 
states of $A$ and the algorithm can be used continuously.
 
Similarly, the time average of any physical quantity $y_i$ --like the
pseudoknot proportion of an RNA molecule-- can be calculated 
by introducing the appropriate {\it time weighted} matrix $\tilde{P}^A\{yt\}$. 
For instance, the time average energy $\bar{E}^A_{ji}$ over all pathways 
between any two states $i$ and $j$ of $A$ is,
$\bar{E}^A_{ji} = {\tilde{P}^A_{ji}\{Et\} / \tilde{P}^A_{ji}\{t\}}$,
where $\tilde{P}^A_{ji}\{Et\}= \sum^{C^A}_{m:j \leftarrow i} \bigl[ \bigl(\sum^{C^A_m} E_ht_h\bigr)\prod^{C^A_m} p_{lk} \bigr]$.

The actual calculation of the probability and path average 
matrices $P^C$ and $\tilde{P}^C$ over a set $C$ of $N$ states will be 
performed recursively in the next subsection. 
As an intermediate step, 
we first
consider hereafter
the {\it unidirectional} connection between two {\it disjoint} 
sets $A$ and $B$. 

Let us hence introduce the transfer matrix
$T^{BA}$ from set $A$ to set $B$ defined as $T^{BA}_{ji}=p_{ji}$, where 
$p_{ji}$ is the probability to make a transition from state $i$ of $A$ to
state $j$ of $B$ ($T^{BA}_{ji}=0$ if $i$ and $j$ are not connected). 
We will assume that $A$ has $n$ states and $B$  $m$ states and that their
probability and path average matrices $P^A$, $\tilde{P}^A$, $P^B$ and $\tilde{P}^B$ 
are known. 
Starting at state $i$ of $A$, we find that the probability to exit on $j$
of $B$ after crossing {\it once and only once} from $A$ to $B$ is,
$p^{eB}_{j}( P^BT^{BA}P^A )_{ji}$,
where we have used matrix notations. Let us consider a particular path
from $i$ in $A$ to $j$ in $B$ crossing {\it once and only once} from 
$A$ to $B$, with statistical weight $\bigl(\prod^Bp_{lk}\bigr)p_{ba}\bigl(\prod^Ap_{l'k'}\bigr)$.
Its contribution to the average time to exit somewhere from the union 
of $A$ and $B$ is,
\begin{eqnarray}
& &\!\!\biggl(\sum_{j \leftarrow b}^Bt_h + \sum_{a \leftarrow
i}^At_{h'}\biggr)\prod_{j \leftarrow b}^Bp_{lk}\cdot p_{ba}\cdot \prod_{a
\leftarrow i}^Ap_{l'k'}= \label{combAandB2}\\
& &\!\!\biggl(\sum_{j \leftarrow b}^B\!t_h \prod_{j \leftarrow b}^Bp_{lk}\!\biggr) p_{ba}
\prod_{a
\leftarrow i}^Ap_{l'k'}\!+\!\prod_{j \leftarrow b}^Bp_{lk} \ p_{ba}
\biggl(\sum_{a
\leftarrow i}^A\!t_{h'}\!\prod_{a
\leftarrow i}^Ap_{l'k'}\!\!\biggr)\nonumber 
\end{eqnarray}
or in matrix form for any ``direct'' pathway from $A$ to $B$,
\begin{equation}
{\cal T}[{P^BT^{BA}P^A}]={\cal T}[{P}^B]\ T^{BA}P^A+P^BT^{BA}{\cal T}[{P}^A], \label{combAandB3}
\end{equation}
which implies that applying the usual differentiation rules to any combination
of probability matrices yields the correct combined path average matrices
(defining ${\cal T}[{T}^{BA}]_{ij}=0$ for all $i$ and $j$).
Note, this out-of-equilibrium calculation of path average quantities is 
reminiscent of the usual equilibrium calculation of thermal averages 
through differentiation of an appropriate Partition Function. Indeed, 
the probability matrices introduced here {\it are} ``partition functions''
over {\it all pathways} within a set of reference states.

\vspace{0.3cm}
\noindent
{\bf The ${\cal O}(n^2)$ algorithm.}
With this result in mind, we can now return to the calculation of the 
probability and path average matrices $P^C$ and $\tilde{P}^C$ for the union 
$C$ of two disjoint sets $A$ and $B$.

Defining $P^{Ab}\!\!=\!P^AT^{AB}\!$ and $P^{Ba}\!\!=\!P^BT^{BA}\!$, 
we readily obtain the probability matrix $P^C$ as an infinite summation 
over {\it all} possible pathway loops between the sets $A$ and $B$ 
(${I}$ is the identity matrix),
\begin{eqnarray}
& P^{C~} =&\left( \begin{array}{c}
            Q^{AA} \, \, \, Q^{AB} \\               
            Q^{BA} \, \, \,  Q^{BB}
\end{array}\right), \,\,\,{\rm with} \label{combAandB5}\\
& Q^{AA} =&\bigl[{I}\!+\!P^{Ab}P^{Ba}\!+\!(P^{Ab}P^{Ba})^2\!+\!\cdots\bigr]P^A\!=\!L^AP^A \nonumber\\
& Q^{BA} =&P^{Ba}L^AP^A  \nonumber\\
& Q^{BB} =&\bigl[{I}\!+\!P^{Ba}P^{Ab}\!+\!(P^{Ba}P^{Ab})^2\!+\!\cdots\bigr]P^B\!=\!L^BP^B \nonumber\\
& Q^{AB} =&P^{Ab}L^BP^B \nonumber
\end{eqnarray}
where $L^A\!=\![{I} - P^{Ab}P^{Ba}]^{-1}$ and $L^B\!=\![{I} - P^{Ba}P^{Ab}]^{-1}$.

Defining also $\tilde{P}^{Ab}=\tilde{P}^AT^{AB}$ and $\tilde{P}^{Ba}=\tilde{P}^BT^{BA}$, 
we finally obtain the path average matrix $\tilde{P}^C$ from simple 
``differentiation'' of the ``partition function'' ${P}^C$, 
Eqs.(\ref{combAandB5}),
\begin{eqnarray}
&\tilde{P}^{C~} =&\left( \begin{array}{c}
            \tilde{Q}^{AA} \, \, \, \tilde{Q}^{AB} \\               
            \tilde{Q}^{BA} \, \, \,  \tilde{Q}^{BB}
\end{array}\right), \,\,\,{\rm with} \label{combAandB6}\\ 
&\tilde{Q}^{AA} =&\tilde{L}^AP^A + L^A\tilde{P}^A \nonumber\\
&\tilde{Q}^{BA} =&\tilde{P}^{Ba}L^AP^A +P^{Ba}\tilde{L}^AP^A +P^{Ba}L^A\tilde{P}^A \nonumber\\
&\tilde{Q}^{BB} =&\tilde{L}^BP^B +L^B\tilde{P}^B  \hspace{3.3cm}\nonumber\\
&\tilde{Q}^{AB} =&\tilde{P}^{Ab}L^BP^B +P^{Ab}\tilde{L}^BP^B +P^{Ab}L^B\tilde{P}^B \nonumber\\  
&\rm where, &\tilde{L}^A =L^A\bigl(\tilde{P}^{Ab}P^{Ba}+P^{Ab}\tilde{P}^{Ba}\bigr)L^A \nonumber\\
& \rm and  &\tilde{L}^B =L^B\bigl(\tilde{P}^{Ba}P^{Bb}+P^{Ba}\tilde{P}^{Ab}\bigr)L^B \nonumber
\end{eqnarray}
Eqs.(\ref{combAandB5}) and (\ref{combAandB6}) are valid for any sizes
$n$ and $m$ of $A$ and $B$. Hence $P^C$ and $\tilde{P}^C$ can be calculated 
recursively starting from $N$ isolated states and $2N$ $1\!\times\!1$ 
matrices $P^i=[1]$ and $\tilde{P}^i\{x\}=[x_i]$, with $i=1,N$, where $x_i$ is 
the value of the feature of interest in state $i$.
Clustering those states 2 by 2, then 4 by 4, etc..., using 
Eqs.(\ref{combAandB5}) and (\ref{combAandB6}) finally yields $P^C$ 
and $\tilde{P}^C$ in ${\cal O}(N^3)$ operations (i.e., by matrix inversions 
and multiplications).
However, instead of recalculating everything back
recursively from scratch each time the set of reference states 
is modified, it turns out to be much more efficient to update it 
continuously each time a single state is added. Indeed,
Eqs.(\ref{combAandB5}) and (\ref{combAandB6}) can be calculated in 
${\cal O}(n^2)$ operations only, when $m=1$ and $n=N\!-\!1$, 
as we will show below.
Naturally, a complete update also requires the removal of one ``old'' 
reference state each time a ``new'' one is added, so as
to keep a stationary number $n$ of reference configurations. As we will
see, this removal step can also be calculated in ${\cal O}(n^2)$ operations 
only.

The ${\cal O}(n^2)$-operation update of the reference set, which we now 
outline, relies on the fact that $T^{AB}$, $P^{Ab}$ and $\tilde{P}^{Ab}$ are 
$n\!\times\!1$ matrices and that $T^{BA}$, $P^{Ba}$ and $\tilde{P}^{Ba}$ are 
$1\!\times\!n$ matrices, when $m=1$ and $n=N\!-\!1$ ($P^B$ and $L^B$ are 
simple $1\!\times\!1$ matrices for a single state $B$). 
Since we operate on {\it vectors}, the Sherman-Morrison 
formula\cite{numrec} can then be used to calculate the $n\!\times\!n$ matrix 
$L^A=\bigl[{I} - P^{Ab}\otimes P^{Ba}\bigr]^{-1}=\bigl[{I} + P^{Ab}\otimes P^{Ba}/(1-P^{Ab}\cdot P^{Ba})\bigr]$. 
Hence, not only
$L^A$ but also any matrix product $L^AM$, where $M$ is a $n\!\times\!n$ 
matrix, can be evaluated in ${\cal O}(n^2)$ operations [by first calculating $P^{Ba}M$ followed by \mbox{$P^{Ab}\otimes (P^{Ba}M)$}]. 
Noticing that the same reasoning applies for the $n\!\times\!n$ matrices 
$\tilde{P}^{Ab}\otimes P^{Ba}$ and $P^{Ab}\otimes \tilde{P}^{Ba}$ provides a simple scheme
to add a single reference state to $A$ and obtain 
matrices $P^C$ and $\tilde{P}^C$ in ${\cal O}(n^2)$ operations using 
Eqs.(\ref{combAandB5}) and (\ref{combAandB6}).

In order to achieve the reverse modification consisting in removing one state
$B$ from the reference set $C$, it is useful to first imagine that the 
original $P^C$ and $\tilde{P}^C$ were obtained by the addition 
of the single state $B$ to the $n$-configuration set $A$, as given by  
Eqs.(\ref{combAandB5}) and (\ref{combAandB6}).
Identifying row $Q^{BA}$, column $Q^{AB}$ and their 
intersection $Q^{BB}$ corresponding to the single state 
$B$ readily yields the vectors $P^{Ab}\!=Q^{AB}/Q^{BB}$, 
$P^{Ba}\!=T^{BA}$ (as $P^B=[1]$) and, hence, the $n\!\times\!n$ matrix
\mbox{$[L^A]^{-1}={I}-P^{Ab}\otimes P^{Ba}=$} ${I}-(Q^{AB}\otimes T^{BA})/ Q^{BB}$.
This gives the following relations between the {\it known}  $L^A$,
 $T^{AB}$, $T^{BA}$, $Q^{AA}$, $Q^{BB}$, $Q^{BA}$, $Q^{AB}$, $\tilde{P}^B$
and $\tilde{Q}^{AA}$, and the {\it unknown} $P^A$ and $\tilde{P}^A$,
\begin{eqnarray}
& Q^{AA} &\!\!=L^AP^A, \nonumber \\ 
& \tilde{Q}^{AA} &\!\!=L^A\!\Bigl[\tilde{P}^A\bigl({I}+T^{AB}\!\otimes Q^{BA}\bigl) +{\tilde{P}^B \over Q^{BB}} Q^{AB}\!\otimes  Q^{BA}\Bigr] \nonumber
\end{eqnarray}
which eventually provides $P^A$ and $\tilde{P}^A$ using the Sherman-Morrison 
formula\cite{numrec} to invert ${I}+T^{AB}\otimes Q^{BA}$,
\begin{eqnarray}
&P^A &\!\!=[L^A]^{-1}Q^{AA}=\Bigl({I}-{Q^{AB}\otimes T^{BA}\over Q^{BB}}\Bigr)Q^{AA}, \label{combAandB9}\\
&\tilde{P}^A&\!\!=\!\Bigl[\!\Bigl(\!{I}\!-{Q^{AB}\!\otimes T^{BA}\over Q^{BB}}\Bigr)\tilde{Q}^{AA}\! -{\tilde{P}^B \over Q^{BB}} Q^{AB}\!\otimes Q^{BA}\Bigr]\!\times \nonumber\\
& &\hspace{0.3 cm}\Bigl({I}-{T^{AB}\otimes Q^{BA} \over 1-T^{AB}\cdot Q^{BA}}\Bigr) \label{combAandB10}
\end{eqnarray}
Hence, the single state $B$ can be removed from the set of reference $C$
in ${\cal O}(n^2)$ operations to yield the updated probability and path average matrices $P^A$ and $\tilde{P}^A$.

Note, however, that this continuous updating procedure, using alternatively
Eqs.(\ref{combAandB5},\ref{combAandB6}) and 
Eqs.(\ref{combAandB9},\ref{combAandB10}) in succession, is expected to become
numerically unstable after too many updates of the reference set. 
For $1\le n \le300$, we have usually
found that the small numerical drifts [as measured e.g. by 
$\epsilon = \sum^A_i(\sum^A_j p^{eA}_j P^A_{ji} - 1)^2 \simeq 0$] 
can simply be reset {every $n^{\rm th}$ update}
by recalculating matrices $P^A$ and $\tilde{P}^A$ recursively from $n$ 
isolated states in ${\cal O}(n^3)$ operations, so as to keep the overall 
${\cal O}(n^2)$-operation count {\it per update} of the reference set.

Another important issue is the choice of the state to be
removed from the updated reference set. Although this choice is 
{\it in principle} arbitrary, the benefit of the algorithm strongly 
hinges on it (for instance removing one of the most statistically 
visited reference states usually ruins the efficiency of the method). 
We have found that a ``good choice'' is {often} the state $j^\star$ 
with the lowest ``exit frequency'' from the current state $i$ [i.e., 
$1/\bar{t}^A_{j^\star i}=\min^A_j\!(1/\bar{t}^A_{ji})$],
but other choices may sometimes prove more appropriate.

\vspace{.5cm}

\noindent
{\large \bf Results}
\vspace{.2cm}

\noindent
{\bf Performance of the ECS algorithm.}
Before applying the ECS algorithm to investigate the prevalence of pseudoknots
in RNA structures, we first focus on the efficacy of the approach 
by studying the net speed-up of the ECS
algorithm with respect to the straightforward algorithm.
As illustrated on Fig~3 for a few natural and artificial sequences, there is 
an {\it actual} $10^1$ to $10^5$-fold increase of the ratio 
``simulated-time over CPU-time'' between ECS and straightforward algorithms 
(black lines) for RNA shorter than 
about $150$~nt, Fig~3.  
This improvement runs parallel to the {\it expected} 
speed-up (grey lines) as predicted by $\bar{\ell}^A_{ji}$, 
Eq.(\ref{meanlengthji}), as long as the number $n$ of reference states 
is not too large (typically $n\le 50$ here), so that the ${\cal O}(n^2)$ 
update routines do not significantly increase the operation count 
as compared to the straightforward algorithm.
\begin{figure}
\includegraphics{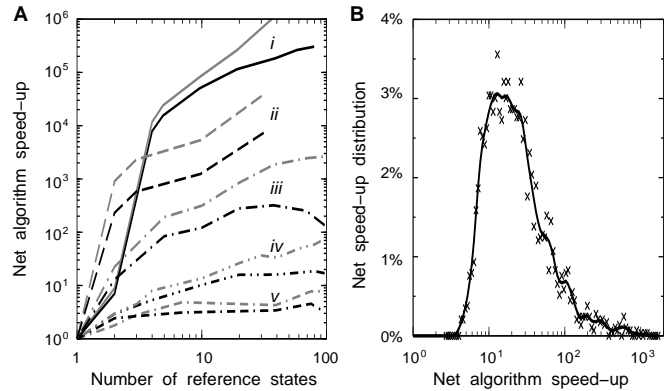}
\caption{\label{fig:epsart}
\small
{\bf A}: 
Expected (grey lines) and actual (black lines)
speed-up of the approach with respect to the 
straightforward algorithm (see main text).
{\bf \em i}: Bistable molecule in Fig~4C (with a combinatorial structure
space of 37 possible helices);
{\bf \em ii}: 67-nt-long molecule with reverse sequence of the bistable molecule
in Fig~4C (38 possible helices). The ${\cal O}(n^2)$ algorithm becomes
unstable above 40 reference states in this case (see main text);
{\bf \em iii}: Hepatitis delta virus ribozyme,  Fig~4B  (84 possible helices);
{\bf \em iv}: average speed-up for random 100-nt-long RNA sequences with 50\% 
G+C~content.
{\bf \em v}: Group I intron ribozyme, Fig~4A  (894 possible helices).
{\bf B}: Net speed-up distribution amongst random 100-nt-long RNA sequences 
with 50\% G+C~content ({\bf \em iv} on Fig~3A) for a cluster of 40 reference 
states.
}
\end{figure}
Hence, the ECS algorithm is most efficient 
for small trapped systems 
(when the dynamics can be appropriately coarse-grained), although 
a several-fold speed-up can still be expected 
with somewhat larger systems, such as the 394-nt-long Group I 
intron pictured in Fig~4A.

Alternatively, using this exact approach may also provide a
controlled scheme to obtain approximate coarse-grained dynamics 
for larger systems.
The C routines of the ECS algorithm 
are freely available upon request.

\begin{figure}
\includegraphics{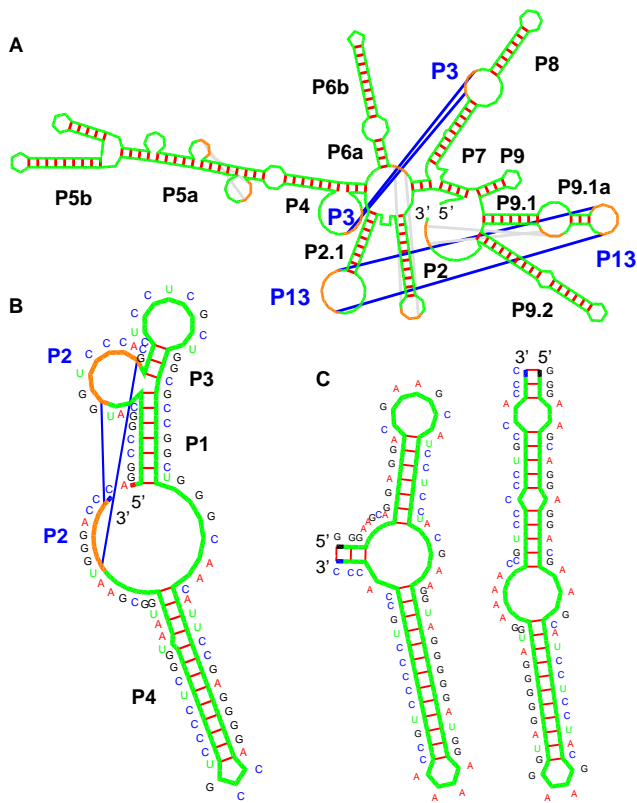}
\caption{\label{fig:epsart} 
\small
RNA structure prediction with the ECS algorithm. 
Structures are drawn using the ``RNAMovies'' software\protect\cite{evers} 
adapted to visualize predicted pseudoknots.
{\bf A} 394-base long
Tetrahymena Group I intron: the lowest free-energy structure found 
shares 80\% base pair identity with the known 3D
structure, including the two main pseudoknots, P3 and 
P13\protect\cite{westhof,williamson1,woodson1,williamson2,woodson2,herschlag}.
{\bf B} 88-base long hepatitis delta virus ribozyme: predicted structure 
shares 93\% base pair identity with the known 3D structure, including the 
main pseudoknot P2\protect\cite{isambert} (but not the 2-base pair long 
P1.1\protect\cite{ferre});
{\bf C} The two structures of a bistable, 67-nt-long artificial RNA molecule.
}
\end{figure}

\vspace{0.3cm}
\noindent
{\bf Pseudoknot prediction and prevalence in RNA structures.}
In the context of RNA folding dynamics, the present approach 
can be used to evaluate time averages for a variety of physical features
of interest, such as the free energy along the folding paths, the fraction 
of time particular helices are formed,
the extension of an RNA molecule 
unfolding under mechanical force\cite{harlepp},
the end-to-end
distance of a nascent RNA molecule during transcription, etc.
Here, we report results on the prediction of pseudoknot prevalence
in RNA structures. They have been obtained performing several thousands of 
stochastic RNA folding simulations 
including pseudoknots. As explained in Theory and Methods, the structural 
constraints between pseudoknot helices and 
unpaired connecting regions are modeled using elementary polymer 
theory (Fig~1C,\cite{isambert}) and added to the traditional base pair 
stacking interactions and simple loops' contributions\cite{turner}. 

We found that many pseudoknots can effectively 
be predicted with such a coarse-grained kinetic approach probing 
seconds to minutes folding time scales.
No optimum ``final'' structure is actually predicted, as such, in this 
folding kinetic approach. 
Instead, low free-energy structures are repeatedly visited, as helices 
stochastically form and break. 
Fig~4A represents the lowest free-energy secondary structure found for  
394-nt-long
Tetrahymena Group I intron, which shows 80\% base pair identity with 
the known 3D structure, including the two main pseudoknots, P3 and 
P13\protect\cite{westhof,williamson1,woodson1,williamson2,woodson2,herschlag}.
A number of smaller known structures with pseudoknots are also compared to
the lowest free-energy structures found with similar stochastic RNA folding 
simulations in\cite{isambert}. In addition, to facilitate the study of 
folding dynamics for specific
RNA sequences, we have set up an online RNA folding server including
pseudoknots at URL {\sf http://kinefold.u-strasbg.fr/}.

\begin{figure}
\includegraphics{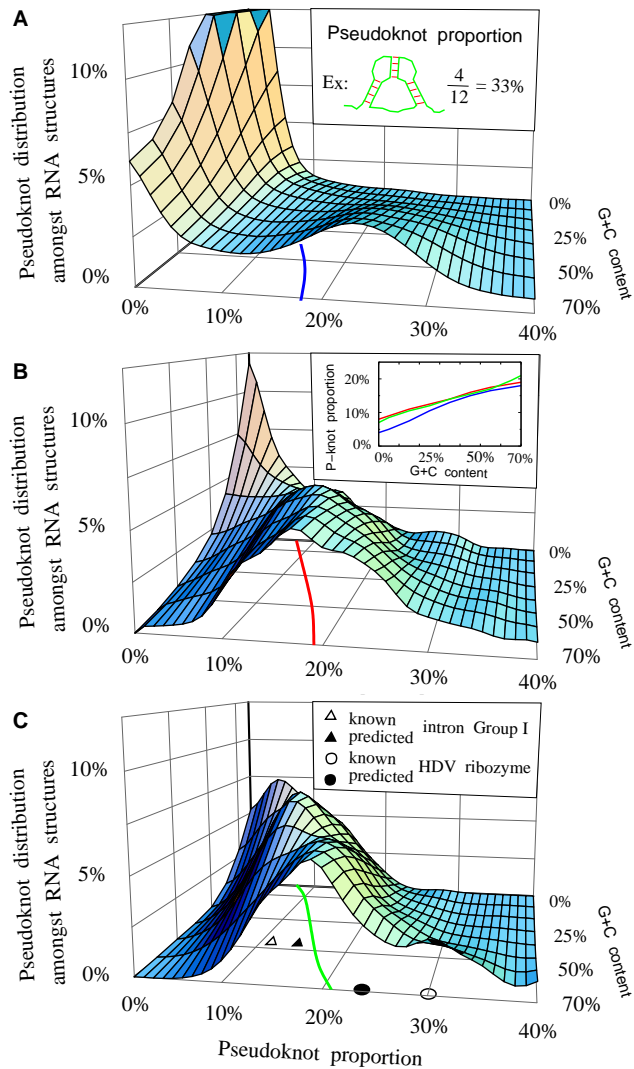}
\caption{\label{fig:dist} 
\small
Distribution of pseudoknot proportion amongst formed 
base pairs for 50-nt-long ({\bf A}), 100-nt-long ({\bf B}), and 150-nt-long 
({\bf C}) random sequences of increasing G+C content.
Projected lines correspond to the average pseudoknot proportion in 50 (blue), 
100 (red), and 150-nt-long (green) random sequences. All three average curves
are displayed in inset on Fig~5B. 
Open (and filled) symbols on Fig~5C correspond to known
(and predicted) pseudoknot proportions for Tetrahymena group I intron, Fig~4A 
(triangles) and Hepatitis delta virus ribozyme, Fig~4B\cite{ferre,isambert} 
(circles).
}
\end{figure}
Beyond specific sequence predictions, we also investigated the general 
prevalence of pseudoknots by studying
 the ``typical'' proportion of pseudoknots 
in both random RNA sequences of increasing G+C content (Fig~5)
and in 150-nt-long mRNA fragments of the {\it Escherichia coli} and 
{\it Saccharomyces cerevisiae} genomes. 
The statistical analysis was done as follows:
for each random and genomic sequence set, 100 to 1000 sequences were sampled
and 3 {\it independent} folding trajectories were simulated for each of them,
using the ECS algorithm.
A minimum duration for each trajectory was determined  
so that more than 80-90\% of sequences visit 
the same free-energy minimum structures along their 3 independent 
trajectories.
The time average proportion of pseudoknots was then evaluated, considering
this fraction of sequences having likely reached equilibrium  
(including the 10-20\% of still unrelaxed sequences does not significantly 
affect global statistics).
In practice, slow folding relaxation limits extensive folding 
statistics to sequences up to 150 bases and 75\%  G+C content, although 
individual folding pathways  can still be studied for molecules up to
250 to 400 bases depending on their specific G+C contents.

The results  for 50-nt-long (Fig~5A), 100-nt-long (Fig~5B), 
and 150-nt-long (Fig~5C) random sequences show, first, a 
{\it broad distribution} in pseudoknot 
proportion from a few percents of base pairs to more than 30\% for some G+C 
rich random sequences. 
Such a range 
is in fact compatible with the various 
pseudoknot contents observed in different known structures ({\it e.g.} see 
triangles and circles in Fig~5C).
Second, the {\it average} proportion of 
pseudoknots (projected curves and inset in Fig~5B) 
slowly increases with G+C content, since stronger (G+C rich) helices
are more likely to compensate for the additional entropic
cost of forming pseudoknots.
Third, and perhaps more surprisingly, this {\it average} proportion of 
pseudoknots appears roughly 
{\it independent of sequence length} except for very 
short sequences with low G+C content (inset in Fig~5B), in  contradiction 
with a naive combinatorial argument.
Fourth, we found that the cooperativity of secondary  
structure rearrangements amplifies the structural consequences of pseudoknot 
formation; typically, a structure with 10 helices including 1 pseudoknot 
conserves {\it not} 9 but {\it only} 7 to 8 of its initial
helices (while 2 to 3 {\it new} nested helices commitantly form)
if the {\it single} pseudoknot is excluded from the structure prediction. 
Thus, neglecting pseudoknots usually induces extended structural 
modifications beyond the sole pseudoknots themselves.

We compared these results with the folding of 150-nt-long sections of 
mRNAs from the genomes of {\it Escherichia coli} (50\% G+C content) and 
{\it Saccharomyces cerevisiae} ({\it yeast}, 40\% G+C content). 
These genomes 
exhibit {\it similar broad distributions of pseudoknots}, 
despites small differences due to G+C content inhomogeneity and
codon bias usage; pseudoknot proportions (mean $\pm$ std-dev.): 
{\it E. coli}, 15.5$\pm$6.5\% (versus 16.5$\pm$7.9\% for 50\% G+C rich random
sequences); 
{\it yeast}, 14$\pm$6.6\% (versus 15$\pm$7.3\% for 40\% G+C rich random
sequences); 
Hence, genomic sequences appear to have maintained a 
large potential for modulating the presence or absence of pseudoknots in 
their 3D structures.

Overall, these results suggest that neglecting pseudoknots in RNA 
structure predictions is probably a stronger impediment than the  
small intrinsic inaccuracy of stacking energy parameters. 
In practice, combining simple structural models (Fig~1C) 
and exactly clustered stochastic (ECS) simulations provides an {\it effective}
approach to predict pseudoknots in RNA structures.

\vspace{.5cm}

\noindent
{\bf Acknowledgements}
\vspace{.2cm}
                   
\noindent
We thank J.~Baschenagel, D.~Evers, D.~Gautheret, R.~Giegerich, W.~Krauth, 
M.~M\'ezard, R.~Penner, E.~Siggia, N.~Socci and E.~Westhof for 
discussions and suggestions. Supported by ACI grants 
n$^\circ$ PC25-01 and 2029 
from Mi\-nist\`ere de la Recherche, France.
H.I. would also like to acknowledge a stimulating two-month 
visit at the Institute for Theoretical Physics, UCSB, Santa Barbara,
where the ideas for this work originated.

\vfill
\eject


\vspace{-.5cm}


\begin{thebibliography}{99}

\bibitem{waterman} Waterman, M.S.
                   (1978) {\em Studies in Found. and Comb.,
                     Adv. in Math. Suppl. Stu.} {\bf 1}, 167-212.

\bibitem{nussinov} Nussinov, R., Pieczenik, G., Griggs, J.R. \& Kleitman D.J. 
                   (1978) {\em SIAM J. Appl. Math.} {\bf 35}, 68-82.

\bibitem{nussinov2} Nussinov, R., \& Jacobson, A.B.  (1980)
                   {\em Proc. Natl. Acad. Sci. USA} {\bf 77}, 
                   7826-7830.

\bibitem{zuker}    Zuker, M. \& Stiegler, P. (1981) {\em Nucleic Acids Res.} 
                   {\bf 9}, 133-148, \hskip 0.2cm and
                   \hskip 0.2cm http://bioinfo.math.rpi.edu/$\sim$mfold/

\bibitem{mccaskill} McCaskill, J.S.  (1990)
                   {\em Biopolymers} {\bf 29}, 
                   1105-1119.

\bibitem{vienna}    Hofacker, I.L. , Fontana, W., Stadler, P.F., Bonhoeffer,
                    L.S., Tacker M. \& Schuster, P.  (1994)
                    {\em Monatsh. Chem.} {\bf 125}, 167-188, \hskip 0.2cm and
                    \hskip 0.2cm http://www.tbi.univie.ac.at/

\bibitem{turner}    Mathews, D.H., Sabina,  J., Zuker, M. \& Turner, D.H.
                    (1999) {\em J. Mol. Biol.} {\bf 288}, 911-940.

\bibitem{higgs}     Higgs,  P.G. (2000) {\em Q. Rev. Biophys.} 
                    {\bf 33}, 199-253, and references therein.

\bibitem{pleij}     Pleij, C.W.A., Rietveld, K., \& Bosch, L.  (1985)
                    {\em Nucleic Acids Res.} {\bf 13}, 1717-1731.
 
\bibitem{tinoco}    Tinoco, I., Jr. (1997)
                    {\em Nucleic Acids Symp Ser.} {\bf 36}, 49-51.

\bibitem{westhof}   Lehnert, V., Jaeger, L., Michel, F. \& Westhof, E. (1996)
                    {\em Chem. Biol.} {\bf 3}, 993-1009.

\bibitem{williamson1}  Zarrinkar, P.P. \& Williamson, J.R.  (1996)
                    {\em Nature Struc. Biol.} {\bf 3}, 432-438.

\bibitem{ferre}     Ferre-D'Amare, A.R., Zhou,  K. \& Doudna, J.A. (1998)
                    {\em Nature} {\bf 395}, 567-574.

\bibitem{woodson1}   Sclavi, B., Sullivan,  M., Chance, M.R., Brenowitz, M. \&
                    Woodson, S.A. (1998) {\em Science} {\bf 279}, 1940-1943.

\bibitem{williamson2} Treiber, D.K., Root,  M.S., Zarrinkar, P.P. \& 
                    Williamson, J.R. (1998) {\em Science} {\bf 279}, 1940-1943.

\bibitem{woodson2}   Pan, J. \& Woodson,  S.A. (1999)
                    {\em J. Mol. Biol.} {\bf 294}, 955-965.

\bibitem{herschlag}  Russell, R., Millet, I.S., Doniach, S. \& Herschlag, D.
                     (2000) {\em Nature Struc. Biol.} {\bf 7}, 367-370.


\bibitem{frameshift} Giedroc, D.P., Theimer, C.A. \& Nixon, P.L.  (2000)
                    {\em J. Mol. Biol.} {\bf 298}, 167-185. Review.

\bibitem{gultyaev}  Gultyaev, A.P., van Batenburg,  E. \& Pleij, C.W.A. (1999)
                    {\em RNA} {\bf 5}, 609-617.

\bibitem{eddy}       Rivas, E. \& Eddy,  S.R.   (1999)
                    {\em J. Mol. Biol.} {\bf 285}, 2053-2068.

\bibitem{isambert} Isambert, H. \& Siggia,  E.  (2000)
                   {\em Proc. Natl. Acad. Sci. USA} {\bf 97}, 6515-6520.

\bibitem{mironov}  Mironov, A.A., Dyakonova,   L.P. \& Kister,  A.E. (1985)
                    {\em J. Biomol. Struct. Dynam.} {\bf 2}, 953-962.

\bibitem{frenkel} Frenkel, D. \& Smit, B. (1996) {\em Understanding Molecular 
Simulation} (Academic  Press) and references therein.

\bibitem{BKL}    Bortz, A.B., Kalos, M.H. \& Lebowitz,  J.L.  (1975)
                 {\em J. Comput. Phys.} {\bf 17}, 10.

\bibitem{mezard}  Krauth, W. \& M\'ezard,  M.  (1995)
                 {\em Z. Phys. B} {\bf 97}, 127.

\bibitem{voter}  Voter, A.F.  (1998)
                 {\em Phys. Rev. B} {\bf 57}, R13985-R13988.
  
\bibitem{pande}  Shirts, M.R. \& Pande, V.S.   (2001)
                 {\em Phys. Rev. Lett.} {\bf 86}, 4983-4987.
  
\bibitem{porschke74bonnet98}
                 P\"orschke, D.  (1974)
                 {\em Biophysical Chemistry} {\bf 1}, 381-386.

\bibitem{krauth} Krauth, W. \& Pluchery, O.   (1994)
                 {\em J. Phys. A; Math. Gen.} {\bf 27}, L715.
  
\bibitem{distribution} In principle, the approach can be adapted to 
                 stochastically drawn lifetimes from known distributions 
                 $P^i(t)$ with mean lifetime $t_i$. This effectively 
                 yields a ${\cal O}(n^3)$ ECS algorithm in this case.

\bibitem{numrec}      Press, W.H., Teukolsky, S.A., Veterling, W.T. \&
  Flannery, B.P.  (1992) {\em Numerical recipes}, 2nd Ed. (University
  Press, Cambridge).

\bibitem{harlepp} Harlepp, S., Marchal, T., Robert, J., L\'eger, J-F.,
  Xayaphoummine, A.,  Isambert, H. and Chatenay, D. (2003)
http://arxiv.org/physics/0309063


\bibitem{evers} Evers, D. \& Giegerich,  R.   (1999)
                    {\em Bioinformatics} {\bf 15}, 32-37.

\end{thebibliography}
\end{document}